\documentclass[traditabstract]{aa} 
\usepackage{graphicx}

\usepackage{txfonts}

\usepackage{natbib}
\newcommand{\bd}{\begin{displaymath}}
\newcommand{\ed}{\end{displaymath}}
\newcommand{\be}{\begin{equation}}
\newcommand{\ee}{\end{equation}}
\newcommand{\beaa}{\begin{eqnarray*}}
\newcommand{\eeaa}{\end{eqnarray*}}
\newcommand{\bea}{\begin{eqnarray}}
\newcommand{\eea}{\end{eqnarray}}

\def\ourlens{SL2S\,J08544$-$0121{}}
\def\HST{{\it HST}{}}

\def\zdP{z_{\rm d,P}}
\def\zdS{z_{\rm d,S}}
\def\zd{z_{\rm d}}
\def\zs{z_{\rm s}}

\def\rtmarg{6.0^{+2.9}_{-2.0} \, {\rm kpc}}
\def\tmarg{1.19\arcsec^{+0.57\arcsec}_{-0.39\arcsec}}
\def\rt{6.0\,{\rm kpc}}
\def\sigmarg{127^{+21}_{-12}{\rm \, km \, s^{-1}}}

\begin{document}
   \title{The Halos of Satellite Galaxies: the Companion of the Massive Elliptical Lens SL2S\,J08544$-$0121\thanks{Based in part on observations made with the NASA/ESA \textit{Hubble Space Telescope}, obtained at the Space Telescope Science Institute, which 
is operated by the Association of Universities for Research in Astronomy, Inc., 
under NASA contract NAS 5-26555. These observations are associated with program 
10876.}}

  \titlerunning{Halo of the Satellite Galaxy in SL2S\,J08544$-$0121}

   \author{S. H. Suyu\inst{1}
          \and
          A. Halkola\inst{2}\inst{,3}
          }

   \institute{Argelander-Institut f\"ur Astronomie, Auf dem H\"ugel 71, 53121 Bonn, Germany\\
              \email{suyu@astro.uni-bonn.de}
         \and
             Excellence Cluster Universe, Technische Universit\"at
             M\"unchen, Boltzmannstr.~2, 85748 Garching, Germany\\
             \email{aleksi.halkola@universe-cluster.de}
          \and
             Tuorla Observatory, Department of Physics \& Astronomy, University of Turku, V\"ais\"al\"antie 20, FI-21500 Piikki\"o, Finland
             }

   \date{Received --; accepted --}

 
  \abstract
  {Strong gravitational lensing by groups or clusters of galaxies
    provides a powerful technique to measure the dark matter
    properties of individual lens galaxies.  We study in detail the mass
    distribution of  
    the satellite lens galaxy in the group-scale lens \ourlens\ by modelling
    simultaneously the spatially extended surface brightness
    distribution of the source galaxy 
    and the lens mass distribution using Markov chain
    Monte Carlo methods.
    In particular, we measure the dark matter halo size of the
    satellite lens galaxy to be $\rtmarg$ with
    a fiducial velocity dispersion of $\sigmarg$.  This is the
    first time the size of an individual galaxy halo in a
    galaxy group has been measured using strong gravitational lensing
    without assumptions of mass following light.  We 
    verify the robustness of our halo size measurement
    using mock data resembling our lens system. 
    Our measurement of the halo size is compatible with the estimated tidal
    radius of the satellite galaxy, suggesting that halos of galaxies
    in groups experience significant tidal stripping, a process
    that has been previously observed on galaxies in clusters.  
    Our mass model of the satellite galaxy is elliptical with its
    major axis misaligned with that of the light by $\sim$$50\deg$.
    The major axis of the total matter distribution is oriented more
    towards the centre of the host halo, exhibiting the radial
    alignment found in \mbox{N-body} simulations and observational
    studies of satellite galaxies.  This misalignment between mass and
    light poses a significant challenge to modified Newtonian dynamics.}
   {}
   {}
   {}
   {}


   \maketitle
%

\section{Introduction}
\label{sec:intro}

Observations to date indicate that the luminous parts
of galaxies are embedded in dark matter halos, but properties of the
dark matter halos are difficult to probe due to the scarcity of
luminous dynamical tracers beyond the visible parts of galaxies
\citep[e.g.,][and references therein]{SofueRubin01, BenderEtal94}.
Gravitational lensing, a phenomenon where foreground mass distributions
deflect the light rays from background source galaxies, provides a powerful
tool to study matter distributions in the Universe because lensing is
sensitive to the total matter distribution independent of the light profiles
of the foreground lenses.  \textit{Strong} gravitational lensing occurs when a
single source galaxy is lensed into multiple images; in contrast, \textit{weak}
gravitational lensing corresponds to weak distortions in the shape of
the single lensed
image of the background source.

Weak galaxy-galaxy lensing has been successfully employed to study the
dark matter halo properties of galaxies both in the field
\citep[e.g.,][]{BrainerdEtal96, dellAntonioTyson96, HudsonEtal98,
  FischerEtal00, SmithEtal01, HoekstraEtal03, HoekstraEtal04, MandelbaumEtal06a, MandelbaumEtal06b, ParkerEtal07, TianEtal09} and in clusters
\citep[e.g.,][]{NatarajanKneib97,NatarajanEtal02, LimousinEtal07, NatarajanEtal09}.  In
this method, the tangential shear signals from the distorted shapes of
background galaxies around foreground galaxies are stacked together to
infer the mass distributions of the foreground lens galaxies.
Stacking is needed because the shear signals around individual
foreground galaxies are typically too weak to constrain the mass
distributions of individual lens galaxies.  Consequently, scaling
relations are used to characterise the average properties of the
foreground lens galaxies in order to model the stacked shear signal.
In the case of galaxy clusters, the cluster profile also needs to be
modelled in detail to extract the residual shear signals around
cluster galaxies \citep[e.g.,][]{LimousinEtal05}. 
From these studies, there is a general trend that
galaxies in denser environments have smaller dark matter halo sizes.

Strong lensing has been used to constrain the size of dark matter
halos in the galaxy cluster Abell 1689 \citep{HalkolaEtal07}.  Scaling
relations are used to relate the halo size to the luminosity of the
galaxy members.  By modelling the positions of the multiple images of
strongly lensed background sources, \citet{HalkolaEtal07} found the
halos of galaxies in Abell 1689 to be truncated compared to galaxies
of equal luminosity in the field.  This is likely due to the tidal
stripping of galaxy halos by the strong gravitational potential of the
galaxy cluster \citep[e.g.,][]{LimousinEtal09b,NatarajanEtal09}.

Both the galaxy-galaxy lensing and strong lensing methods mentioned above are
statistical in that they average signals over many lens galaxies and
study the ensemble properties of these lens galaxies.  Recently, two studies
of strong lensing clusters have yielded halo size measurements for individual
cluster galaxies.  \citet{RichardEtal10} modelled the cluster Abell 370 and
used the multiple image positions to constrain the halo size of a cluster
galaxy that lies in the vicinity of a giant gravitationally lensed arc.
\citet{DonnarummaEtal10} analysed the cluster Abell 611 using the multiple
image positions of lensed arcs, and measured the halo sizes of the brightest
cluster galaxy (BCG) and several galaxies in the vicinity of the BCG.  In both
studies, the centroids, axis ratios and position angles of the cluster galaxies
have been fixed to the observed light distributions.
While studies such as the Sloan Lens ACS Survey indicate that
light is generally a good tracer of mass for isolated galaxy
lenses in the field \citep{KoopmansEtal06, 
BoltonEtal08a, BoltonEtal08b}, there can be misalignments especially
for galaxies in dynamical environments such as in groups and clusters \citep[e.g.,][]{PereiraEtal08}.
If there are misalignments in the cluster galaxies in Abell 370 and Abell 611, then fixing
these lens parameter values to the observed light distribution could lead to a
bias in the determination of the 
halo size.  Allowing all parameters to vary, while avoiding the bias, would
in general induce parameter degeneracies.

In this paper, we aim to constrain the halo size of the satellite
galaxy of the massive elliptical lens in \ourlens.  We show how we can
use strong lensing to constrain individual galaxy halo size
\textit{without assumptions of scaling relations and without
enforcement of mass following light}.  This technique works on
group- and cluster-scale lens systems where  lensed arcs 
are comparable to, or larger than, the
halo sizes of
group/cluster galaxies.  In lens systems where a group/cluster galaxy
lies close to an arc, the morphological information of the arc
provides powerful constraints on the mass distribution of the galaxy.
Recently, \citet{VegettiEtal10} studied the system SDSS
J120602.09+514229.5, which exhibits
similar lensing features as in \ourlens.  They used the lensing arc to
detect the presence of mass substructure (a visible dwarf satellite
galaxy in this case), which demonstrates their dark matter
substructure detection method \citep{VegettiKoopmans09}.  Properties of
the dwarf satellite such as its mass were measured, nonetheless, the
halo size of the dwarf galaxy was assumed to be its theoretical tidal radius.  Our
aim in studying \ourlens\ is to measure the halo size
of the satellite galaxy.

The paper is organised as follows.  We summarise
the observations of the lens system \ourlens\ in Section
\ref{sec:obs}, and describe the reduction of the observations in
Section \ref{sec:hstAnalysis}.  We present the lens modelling method
and the measurement of the satellite halo size in Section
\ref{sec:LensModel}.  Simulations that test this method for the case of
\ourlens\ are in
Section \ref{sec:simtest}.  Finally, we discuss our results in Section
\ref{sec:discuss} before concluding in Section \ref{sec:conclude}.

Throughout the paper, parameter constraints are given by the median
values with the uncertainties given by the 2.3 and the 97.7
percentiles (corresponding to 95.4\% credible intervals (CI)) of the
marginalised probability density distributions.  We assume a flat
$\Lambda$-CDM cosmology with $H_0=70 \rm{\, km\, s^{-1}\, Mpc^{-1}}$ and
$\Omega_{\Lambda}=1-\Omega_{\rm M}=0.72$.  From the redshifts of the lens and the
source galaxies in Section \ref{sec:obs}, one arcsecond at the lens (source) 
plane in \ourlens\ corresponds to $5.0\ (8.5) {\rm \, kpc}$.  The position
angles of the galaxies are measured eastward from north.


\section{Observations of \ourlens}
\label{sec:obs}

The lens system \ourlens, discovered in the Canada France Hawaii
Telescope Legacy Survey (CFHTLS), is part of the Strong Lensing Legacy
Survey (SL2S) \citep{CabanacEtal07, LimousinEtal09a}.  Details of the 
observations of the lens system are described in \citet{LimousinEtal09a} and
\citet{LimousinEtal09c}; we summarise below the relevant ones for this paper.

\begin{figure}
  \includegraphics[width=1.0\columnwidth,bb=20 20 340 290]{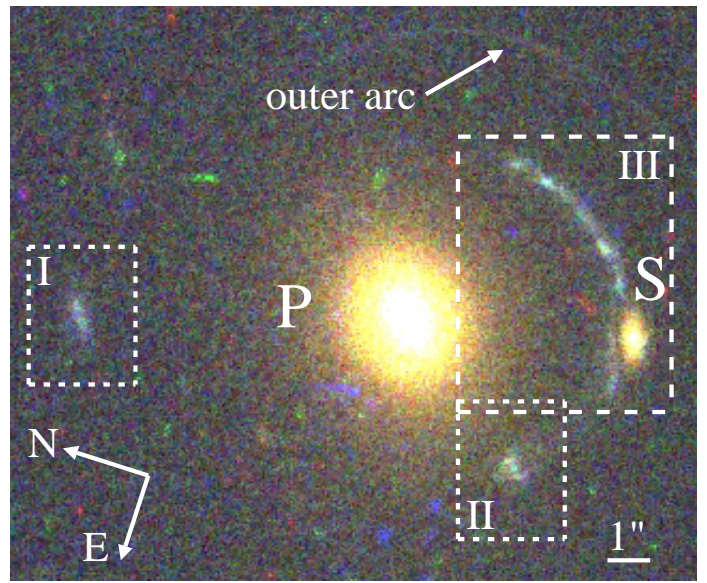}
  \caption{A colour image of \ourlens\ constructed from the
    F475W, F606W and the F814W images taken with the \HST/ACS. The primary
    and satellite lenses are denoted by P and S, respectively. 
    The boxes mark the multiple image
    systems that are shown in detail in Fig.~\ref{fig:hstPtImMarked}.}
  \label{fig:hstcolor}
\end{figure}

\textit{Hubble Space Telescope} (\HST) Advanced Camera for Surveys
(ACS) images were obtained for this lens system in snapshot mode in
three filters: F475W, F606W and F814W (ID 10876; PI Kneib).  
Figure \ref{fig:hstcolor} shows a massive galaxy, which we refer to as the
primary lens galaxy (labelled P), that strongly lenses two
background source galaxies: one into
a bright inner arc (with its corresponding images, shown in the 
marked boxes) and the other into a faint
outer arc (as marked on the figure).  In this 
work, we focus on the bright inner arc because the outer arc is
too faint for (i) redshift measurements and (ii) providing useful
information on the system in addition to the inner arc without
redshift information.  Discussions
of the arc hereafter refer to the inner arc unless otherwise
stated.  The arc is at a distance of $\sim$$5\arcsec$ from the
primary lens galaxy, making this a group-scale lens system.  This system is
special in that a galaxy, labelled S, lies near the arc.  

The
redshifts of the primary lens and the source are
$\zdP=0.3530\pm0.0005$ and $\zs=1.2680\pm0.0003$, respectively
\citep{LimousinEtal09a, LimousinEtal09c}.  The redshift of galaxy S is
$\zdS=0.3514\pm0.0004$ (Mu\~{n}oz et al., in preparation).  The
uncertainties in the redshift measurements are statistical only; systematic
errors due to, e.g., wavelength calibration and dependence on section of
spectrum for redshift extraction, are $\sim$$0.0005$ (R.~Cabanac 2010,
priv.~comm.).  We note that a difference
of $\Delta z=0.001$ between galaxy P and galaxy S corresponds to
a line-of-sight velocity difference of $\sim$$200 {\rm\, km\, s^{-1}}$ at $\zdP$,
which is consistent with the expected radial velocities in a galaxy group.
Based on the proximity of galaxy S to galaxy P and their measured $\zdP$ and $\zdS$, we interpret galaxy S as being
physically associated with the primary lens galaxy, and we refer to
galaxy S as the satellite lens galaxy.  For the analysis, we use $\zd$=$\zdP$
as the lens redshift for the system.  

The positions of the multiple images of the background source show a
high degree of asymmetry: the northern image is $\sim$$8\arcsec$ away from
the primary lens galaxy, significantly larger than the
$\sim$$5\arcsec$ of the arc.  This is caused by the group environment of the
lens system: a concentration of galaxies is
located about $\sim$$1\arcmin$ to the east of the system and has
similar redshifts as the lens system.  
\citet{LimousinEtal09c} showed that this asymmetry can be exploited
to study the global properties of the group.

\section{\HST\ Image Analysis}
\label{sec:hstAnalysis}

We obtain the standard CALACS pipeline\footnote{A pipeline for
  calibrating ACS data, developed at the Space Telescope Science
  Institute} reduced images from the 
ESO/ST-ECF\footnote{The European Southern Observatory (ESO), and the
  Space Telescope European Coordinating Facility (ST-ECF)} Science Archive. Since single exposures were taken in each
wavelength band, these
images still contain cosmic rays that are subsequently masked using
L.A.Cosmic \citep{vanDokkum01}. The data are further reduced using the
MultiDrizzle package for IRAF \citep{FruchterEtal09} to correct for the
geometric 
distortion.  The output image pixel size is $0.05\arcsec$. 

For the reconstruction of the lensed arcs, it is important to remove
any light contribution to the arcs from the lens galaxies. This is
done using the GALFIT software package \citep{PengEtal02}. 
Following, e.g., \citet{MarshallEtal07} and \citet{SuyuEtal09}, the point
spread function (PSF) is estimated from a star in the field.
The regions around 
the lensed arcs, as well as those of the remaining cosmic rays, are
masked for extracting the lens galaxy light.  The light
distribution of the primary lens is bimodal with a concentrated
circular component and a smoother, more elliptical component. The 
satellite lens and the two
components of the primary lens are fitted
simultaneously with S\'{e}rsic profiles \citep{Sersic68}.   To 
investigate the light distribution of the primary lens as a whole, we
also fit a single S\'{e}rsic profile to the primary lens. In
doing so, we mask out the core of the primary lens that shows clear
bimodality in the light distribution. 
We set the origin of the image coordinate system to be located at the
centroid of the single-component primary lens galaxy.  The best fit
values for the S\'{e}rsic profiles are tabulated in Table
\ref{tab:galfit}. The profile parameters for the satellite lens are
nearly the same in both the single- and two-component modelling of the
primary lens light.

\begin{table}
  \caption{Properties of the lens galaxy light distribution
  }
  \label{tab:galfit}
  \begin{center}

  \begin{tabular}{lcccccc}
    \hline
    Component & $x$    & $y$    & $R_{e}$ & $n$ & $q$ & $\phi$ \\
              & ($\arcsec$) & ($\arcsec$) &     ($\arcsec$)  &    &      &  (\degr) \\
    \hline\hline
    Primary 1 &    $-0.06$  &     $\phantom{-} 0.21$  &       2.55  &     5.74  &
    0.95  &    114 \\ 
    Primary 2 &    $\phantom{-} 0.09$  &    $-0.21$  &       1.53  &     2.91  &
    0.71  &   137 \\ 
    Satellite &    $\phantom{-} 5.27$  &    $-0.45$  &       0.41  &     2.33  &
    0.47  &    115 \\
    \hline 
    Primary   &     $\equiv$0.00  &     $\equiv$0.00  &     2.80  &
    7.97  &     0.79  &   131 \\
    Satellite    &   $\phantom{-} 5.27$  &    $-0.45$  &       0.41  &     2.34  &
    0.47  &    115\\ 
    \hline
  \end{tabular}
  \end{center}
  Notes -- The top (bottom) half gives S\'{e}rsic profile parameters
  from GALFIT for a two (single) component primary lens.  
  Columns 1 and 2 denote the coordinate of the
  centroid, column 3 is the effective radius, column 4 is the S\'{e}rsic
  index, column 5 is the axis ratio, and column 6 is the position angle.

\end{table}

Despite the dwarfish appearance of the satellite galaxy
  when juxtaposed with the primary lens, the satellite galaxy is a normal
elliptical galaxy based on its effective radius in Table
\ref{tab:galfit} and the lensing derived velocity dispersion in
Section \ref{sec:discuss:lit} \citep{TollerudEtal10, GravesEtal09,
  GehaEtal03}.

We have chosen to use the F606W image for our analysis since this has
the highest signal-to-noise ratio (SNR) and is least affected by cosmic
rays near the lensed arcs.

\section{Satellite Halo Size from Lens Modelling}
\label{sec:LensModel}

In this section, we describe our lens modelling of the \HST\ data for
constraining the halo size of the satellite galaxy.  We model the lens
system using simply-parametrised mass profiles (Section
\ref{sec:LensModel:profiles}), and sample the posterior probability
distribution of the lens parameters using Markov chain Monte Carlo
(MCMC) methods (Section \ref{sec:LensModel:MCMC}).  To constrain the
lens parameters, we use either the image positions of the multiply
imaged source (Section \ref{sec:LensModel:ImPos}) or the extended
surface brightness distribution of the lensed source galaxy (Section
\ref{sec:LensModel:ExtIm}).  This allows us to quantify the amount of
additional information the extended images provide on the mass
distribution of the lenses.

\subsection{Simply-parametrised lens profiles}
\label{sec:LensModel:profiles}
For a review on gravitational lensing, we refer the reader to
\citet{SchneiderEtal06}. 
We describe the mass distribution for the lens system as two lens galaxies
(the primary and the satellite) in the presence of a constant external shear.

The primary lens galaxy is modelled as a singular pseudoisothermal elliptic mass
distribution \citep[PIEMD;][]{KassiolaKovner93} with dimensionless
surface mass density
\be
\label{eq:piemd}
\kappa_{\rm P}(\theta_1, \theta_2) = \frac{b_{\rm P}}{2\theta_{\rm em}},
\ee
where 
\be
\label{eq:rem}
\theta_{\rm em}^2 =\frac{{\theta_1^2}}{(1+\epsilon)^2}+\frac{\theta_2^2}{(1-\epsilon)^2}, 
\ee
$(\theta_1,\theta_2)$ are angular coordinates on the image/lens plane,
and $\epsilon$ is the ellipticity
defined as $\epsilon\equiv(1-q)/(1+q)$ with $q$ being the axis ratio.  The
strength and axis ratio of the primary lens are $b_{\rm P}$ and $q_{\rm P}$,
respectively.
The distribution is appropriately translated by the centroid position
($\theta_{\rm 1,P}$, $\theta_{\rm 2,P}$) and rotated by the position angle
$\phi_{\rm P}$. 

The satellite lens
galaxy is modelled as a dual pseudo isothermal elliptical mass distribution
\citep[dPIE;][]{EliasdottirEtal07} with a vanishing core radius,
\be
\label{eq:dpie}
\kappa_{\rm S}(\theta_1,\theta_2) = \frac{b_{\rm S}}{2}\left(\frac{1}{\theta_{\rm em}} - \frac{1}{\sqrt{\theta_{\rm em}^2+t^2}} \right),
\ee
where 
$b_{\rm S}$ is the satellite lens strength, $\theta_{\rm em}$ is
defined in equation (\ref{eq:rem}) with $q_{\rm S}$ being the axis
ratio of the satellite, and $t$ is the ``truncation radius'' of the
satellite.  The 
dPIE distribution is suitably translated by the centroid position
($\theta_{\rm 1,S}$, $\theta_{\rm 2,S}$) and rotated by the position angle
$\phi_{\rm S}$.  The
radial dependence of the three dimensional mass density distribution
corresponding to equation (\ref{eq:dpie}) is 
\be
\label{eq:dpie_rho}
\rho(r) \propto \frac{1}{r^2(r^2+r_{\rm t}^2)},
\ee
where $r$ is the three dimensional radius, $r_{\rm t}=D_{\rm
    d}t$, and $D_{\rm d}$ is the angular diameter 
distance between the observer and the lens.
Consequently, for $r \ll r_{\rm t}$ the mass
distribution of the satellite galaxy is isothermal ($\rho \propto
r^{-2}$), but for $r \gg r_{\rm t}$ the mass distribution falls
off as $r^{-4}$.  We refer to $r_{\rm t}$ as the size of
the satellite dark matter halo (with the corresponding angular size as
$t$), but note that $r_{\rm t}$ is roughly the
half-mass radius \citep[e.g.,][]{EliasdottirEtal07}.  The fiducial
velocity dispersion is defined as  
\be
\label{eq:sigmadpie}
\sigma_{\rm dPIE} = c \left( \frac{b_{\rm S}}{6\pi}\frac{D_{\rm
      s}}{D_{\rm ds}}\right)^{1/2},
\ee
where $c$ is the speed of light, and $D_{\rm s}$ ($D_{\rm ds}$) is the
angular diameter distance between the observer (lens) and the source.
The appendix in \citet{EliasdottirEtal07} shows the relation between
$\sigma_{\rm dPIE}$ and the measured velocity dispersion
$\sigma_{*}$. 

The constant external shear is parametrised by a shear strength
$\gamma_{\rm ext}$ and a shear angle $\phi_{\rm ext}$.  In terms of
polar coordinates $\vartheta$ and $\varphi$ such that
$\theta_1=\vartheta \cos(\varphi)$ and $\theta_2=\vartheta
\sin(\varphi)$, the lens potential describing the constant external
shear is
\be
\label{eq:extsh}
\psi_{\rm ext}(\vartheta,\varphi)=\frac{1}{2} \gamma_{\rm ext}
\vartheta^2 \cos(2(\varphi-\phi_{\rm ext})),
\ee
where the shear centre is
arbitrary since it corresponds to an unobservable constant shift in the source
plane.  Note that $\kappa_{\rm ext}=\frac{1}{2}
\nabla^2 \psi_{\rm ext}$ is zero.
The shear position angle of $\phi_{\rm ext}=0\degr$ or $\phi_{\rm
  ext}=90\degr$ corresponds to a shearing along the
$\theta_1$-direction or the $\theta_2$-direction, respectively.

We have 13 lens parameters in total: 5 for the PIEMD ($\theta_{\rm 1,P}$,
$\theta_{\rm 2,P}$, $q_{\rm P}$, $\phi_{\rm P}$, $b_{\rm P}$), 6 for
the dPIE ($\theta_{\rm 1,S}$, $\theta_{\rm 2,S}$, $q_{\rm S}$,
$\phi_{\rm S}$, $b_{\rm S}$, $t$) and 2 for the external shear
($\gamma_{\rm ext}$, $\phi_{\rm ext}$).  
We impose uniform priors on all lens parameters.  However, for $t$, we
restrict its range to be between $0.4\arcsec$ and $10\arcsec$.  The
lower limit is set by the effective radius of the satellite galaxy
(the radius within which the integrated flux of the galaxy is half its
total flux) in Table \ref{tab:galfit}.  Any dark matter contribution
to the satellite galaxy would lead to a value for $t$ (approximately
the half-mass radius) that is larger than the effective radius,
assuming a roughly constant mass-to-light ratio for the luminous
matter. The upper limit of $10\arcsec$ is based on the expectation
that a satellite halo size larger than the arc ($\sim$$5\arcsec$) 
will not be constrained.

\subsection{MCMC sampling}
\label{sec:LensModel:MCMC}

We follow \citet{DunkleyEtal05} to achieve efficient MCMC sampling of
the posterior probability density function (PDF) of the lens
parameters and to test for convergence.  The PDFs for the lens
parameters are described in Sections \ref{sec:LensModel:ImPos} and
\ref{sec:LensModel:ExtIm}.  Efficiency of the MCMC sampling hinges upon the
proposal density distribution (which encodes the information regarding
the direction and step size for obtaining the next random step in the
MCMC chain).  Briefly, the procedure is: (i) obtain a chain with
acceptance rate of $\sim$$0.25$ based on an initial guess of the
proposal density distribution; (ii) update the proposal density
distribution by approximating it as a multivariate Gaussian with its
covariance matrix set to the covariances among the parameters of the
previous chain after the burn-in phase;  (iii) adjust the Gaussian
widths of the proposal density distribution by a constant factor to
obtain a new chain with an acceptance rate of $\sim$$0.25$: a global
increase (decrease) in the Gaussian widths of the proposal density
distribution leads to a lower (higher) acceptance rate;
(iv) repeat (ii) and (iii) until a convergent
MCMC chain is obtained.  We find that the proposal
density distribution needs updating only a few times to achieve efficient
sampling of the posterior PDF.  We follow the
power-spectrum convergence test introduced in \citet{DunkleyEtal05}.
In this approach, convergence is achieved when the chain is drawing
samples throughout the region of high probability (so that the
correlation between successive samples in the chain does not bias the
posterior PDF) and when the ratio of the variance of the sample mean
to the variance of the sample distribution is $\leq 0.01$ (so that statistics
can be obtain with good accuracy).

\subsection{Lens modelling using image positions}
\label{sec:LensModel:ImPos}
\begin{figure}
  \includegraphics[width=1.0\columnwidth]{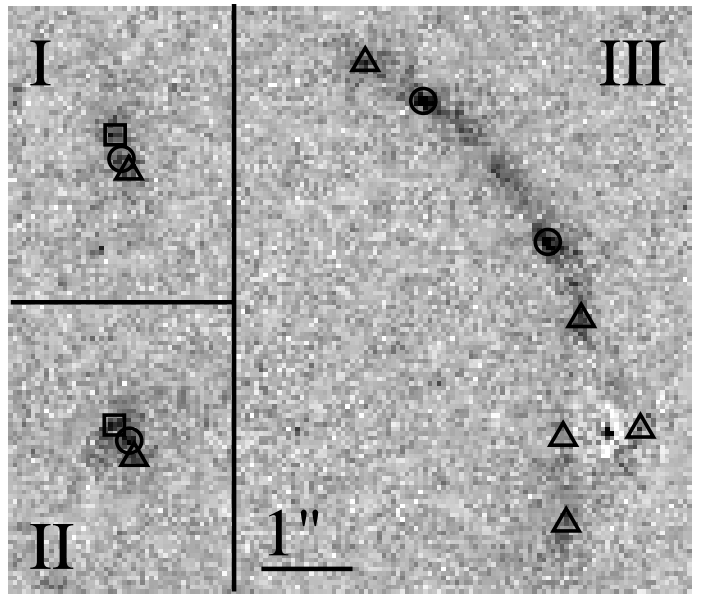}
  \caption{Identification of the multiple images. The three panels (I,
    II and III) correspond to the regions marked in Fig.~\ref{fig:hstcolor}.  
    These image cutouts are taken from the F606W
    image with the light from the lens galaxies subtracted. The
    triangles, circles and squares show the three multiple image systems.}
  \label{fig:hstPtImMarked}
\end{figure}

We identify three peaks in the source surface brightness distribution
in the \HST\ F606W image.  Figure \ref{fig:hstPtImMarked} shows the
identified image positions corresponding to each of the three peaks in
the source.  The three peaks are strongly lensed into 2, 4 and 7
images, respectively.  We include the images around the satellite
galaxy which are important for our study of the satellite. These
images were not considered by \citet{LimousinEtal09c} since the focus
of their study was to probe the global properties of the galaxy group
instead of local effects due to the satellite.  We estimate the
uncertainties in the 
image positions to be $0.05\arcsec$, except for the image positions near
the satellite galaxy whose uncertainties we take to be $0.15\arcsec$
as they are more difficult to identify.  The image positions provide a
total of 26 constraints.

We follow \citet{HalkolaEtal06, HalkolaEtal08} for the lens
modelling based on image positions.  The posterior PDF for the lens
parameters, collectively denoted as $\vec{\eta}$, given the image
positions $\vec{d}_{\rm pt}$ is 
\be
\label{eq:post_pt}
P(\vec{\eta} | \vec{d}_{\rm pt}) \propto \overbrace{P(\vec{d}_{\rm pt} |
\vec{\eta})}^{\rm likelihood} \overbrace{P(\vec{\eta})}^{\rm prior}.
\ee
The likelihood is 
\be
\label{eq:like_pt}
P(\vec{d}_{\rm pt} |\vec{\eta}) = \frac{1}{Z_{\rm {pt}}} \exp
  {\left[-\frac{1}{2}\displaystyle\sum_{j=1}^{N_{\rm
          sys}}\displaystyle\sum_{i=1}^{N_{\rm im}^j}
      \frac{\vert\vec{\theta}_{i,j}-\vec{\theta}_{i,j}^{\rm
          pred}(\vec{\eta})\vert^2}{\sigma_{i,j}^2} \right]},
\ee
where $N_{\rm sys}$ is the number of multiply imaged systems (=3 in
our case, corresponding to the 3 peaks in the source surface brightness
distribution), $N_{\rm im}^j$ is the number of multiple images in system
$j$, $\vec{\theta}_{i,j}$ is the observed image position, 
$\vec{\theta}_{i,j}^{\rm pred}(\vec{\eta})$ is the modelled image
position, $\sigma_{i,j}$ is the uncertainty in the observed image
position, and $Z_{\rm pt}$ is the normalisation given by
\be
\label{eq:like_pt_norm}
Z_{\rm pt}={(2\pi)^{N_{\rm pt}/2}
  \displaystyle\prod_{j=1}^{N_{\rm sys}} \displaystyle\prod_{i=1}^{N_{\rm im}^j} \sigma_{i,j}}
\ee
with
\be
N_{\rm pt}=\displaystyle\sum_{j=1}^{N_{\rm sys}} {N_{\rm im}^j}.
\ee
Positions of the surface brightness peaks in the source plane,
  $\vec{\beta}_{j}$ (where $j=1\ldots N_{\rm sys}$), are needed to
  predict the image positions $\vec{\theta}_{i,j}^{\rm
    pred}(\vec{\eta})$.  For a given set of values for the 13 lens
  parameters $\vec{\eta}$, we optimise for the source
  positions, denoted by $\vec{\hat{\beta}}_{j}$, that maximise
  equation (\ref{eq:like_pt}).  Effectively, we have marginalised over
  the source position parameters by approximating the likelihood as
  having a delta
  function at $\hat{\vec{\beta}}_{j}$.
As described in Section \ref{sec:LensModel:profiles}, the prior PDF,
$P(\vec{\eta})$, is uniform with $t$ limited to $[0.4\arcsec,10\arcsec]$.

\begin{figure}
\centering
  \includegraphics[width=1.0\columnwidth,bb=18 160 592 640]{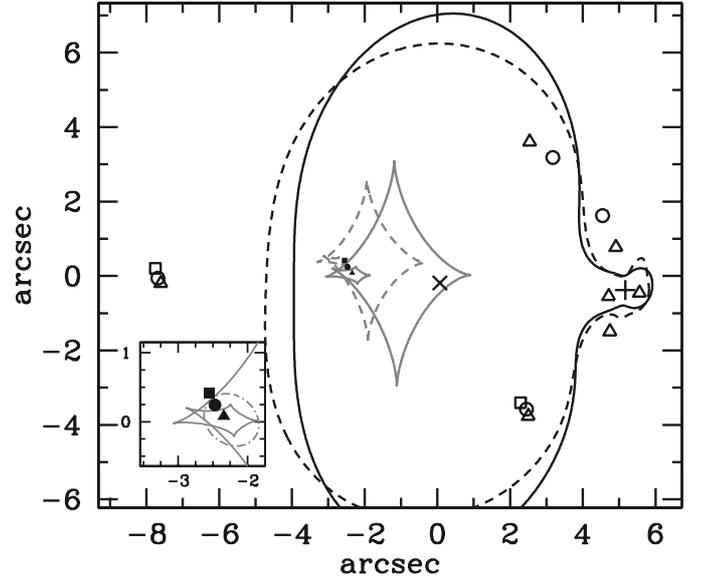}
\caption{\label{fig:ImPosCritCaus}  Critical curves (black) and
  caustics (grey) of the lens 
  models.  Dashed (solid) curves correspond to the most probable lens model
  with image positions (extended image surface brightness) as constraints.
  The dot-dashed curve in the inset marks the
  cut of the satellite.
  The cross (plus) symbol denotes the modelled centroid position of
  the primary (satellite) lens.
  The three types of open symbols mark the three sets of multiple
  image systems.  The solid symbols are the corresponding
  modelled source positions based on the most probable lens parameters from
  the extended image modelling.  The shapes of the critical and caustic
  curves curves are similar for the two models, but
  the model based on image positions (dashed) has a rounder primary lens with
  higher external shear. }
\end{figure}

We sample the posterior PDF $P(\vec{\eta} | \vec{d}_{\rm pt})$ based
on the MCMC procedure described in Section \ref{sec:LensModel:MCMC}.
The critical curves and caustics
of the most probable lens model and the source
positions are shown in Fig.~\ref{fig:ImPosCritCaus}.  The large-scale
diamond-shaped caustic curve, the astroid, is associated with the primary
lens.  
The presence of the 
satellite galaxy transforms an 
ordinary cusp of the astroid into a pentagon-shaped caustic curve.  Due to the
singular nature of the satellite mass profile, there is also a
``cut'' (shown as a dot-dashed curve in the inset), which is the limiting case of a
caustic curve \citep[e.g.,][]{KormannEtal94}.  
Each of the three source surface brightness peaks lies in a different
caustic region, and is thus multiply imaged different number of times.
The source peak marked by the solid square lies outside of the main
astroid caustic of the primary lens, and hence has 2 images.  The solid
circle is inside both the astroid caustic and the cut, and therefore has
a total of 5 predicted images (the caustic crossing creates 2
additional images and the cut crossing yields 1 more).  We have
observed only 4 image positions (marked by open circles in
Fig.~\ref{fig:ImPosCritCaus}) because the $5^{\rm th}$ predicted image, which
lies near the centre of the satellite and is less magnified than the other images, could not be
detected given the noise levels and the residuals of the lens
galaxy light near the centre (see Fig.~\ref{fig:hstPtImMarked}).  The
source marked by the solid triangle, being inside the pentagon caustic, has two
more predicted images than the solid circle and is thus a 7-image
system.  The creation of such a 7-image system is not possible without
the satellite galaxy -- the primary elliptical lens galaxy with its 
simple astroid can only produce 4 images (excluding the central, often
demagnified, image that would lie near the centre of the primary
lens).  All in all, we are able to reproduce all the observed image
positions without predicting extra images that can be detected in the
data.  The root mean square separation between the observed and the
corresponding predicted
image positions of the most probable lens model is $0.051\arcsec$.
  
\begin{figure}
\centering
\includegraphics[width=1.0\columnwidth,bb=45 170 595 705]{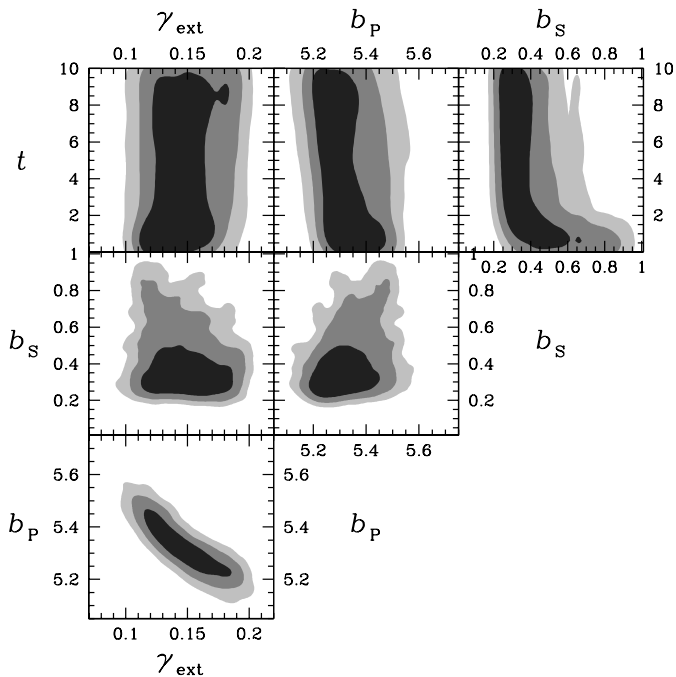}
\caption{\label{fig:ImPosMCMC} Marginalised posterior PDF for $\gamma_{\rm
    ext}$ (external shear strength), $b_{\rm P}$ (strength of primary
  lens in arcseconds), $b_{\rm S}$ (strength of satellite lens in arcseconds)
  and $t$ (truncation radius of satellite in arcseconds) 
  based on image position modelling.  The three
  shaded areas show the 68.3\%, 95.4\% and 99.7\% credible
  regions. The halo size of the satellite galaxy is not constrained
  using image positions.}
\end{figure}

The results of the MCMC sampling for a subset of the 13 parameters are
shown in Fig.~\ref{fig:ImPosMCMC}.
There is a degeneracy between $b_{\rm P}$ and $\gamma_{\rm ext}$: a
higher lens strength produces more internal shear so that less
external shear is needed to explain the configuration of the multiple
images.  Similarly, there is a correlation between $q_{\rm P}$  (not
shown) and $\gamma_{\rm ext}$: a rounder primary lens (i.e., higher
$q_{\rm P}$) with less internal shear requires more external shear.  
The figure shows that both $b_{\rm S}$
and $t$ are not well constrained.  In particular, $t$ shows a full
degeneracy in the prior range of $0.4\arcsec$ and $10\arcsec$.  The L-shaped
contours of $t$ and $b_{\rm S}$ indicate that the posterior cannot be
well described by a multivariate Gaussian distribution.  As a result, the
multivariate Gaussian proposal density distribution is suboptimal and
the efficiency of the MCMC sampling is hampered by these parameter
degeneracies.  In fact, we could not obtain a convergent chain with
$>2\times10^6$ 
samples even after several updates of the proposal density
distribution due to the parameter degeneracies.  The results
presented in this section are based on a chain of length $10^6$ after
the burn-in phase.

Table \ref{tab:LensParaRealData} lists the marginalised lens parameters in the
second column.  The value of $t$, which nearly spans the entire range of the
prior, shows that we cannot constrain the halo
size of the satellite galaxy using image positions as constraints.
The remedy is to use more information, such as the surface brightness
of each pixel in the extended lensed images that is described next.

\begin{table}
\caption{Modelled Lens Parameter Values}             
\label{tab:LensParaRealData}      
\centering                          
\renewcommand{\arraystretch}{1.4}  
\begin{tabular}{lcc}        
\hline                 
\multicolumn{1}{l}{Parameter} & \multicolumn{2}{c}{Marginalised value (95.4\% CI)} \\    
\multicolumn{1}{l}{}          & Positions & Extended \\
\hline                        
\hline
 $\phantom{ }\gamma_{\rm ext}$                      & $  \phantom{-}0.15_{- 0.03}^{+ 0.04}$ &  $ \phantom{-}0.11_{-0.01}^{+0.01}$  \\
 $\phantom{ }\phi_{\rm ext}$            $[\degr]$   & $  26_{- 4}^{+6}$ &  $ 29_{-2}^{+2}$  \\
 $\phantom{ }\theta_{1,{\rm P}}$        $[\arcsec]$ & $   -0.13_{- 0.26}^{+ 0.18}$ &  $ \phantom{-}0.08_{-0.07}^{+0.08}$  \\
 $\phantom{ }\theta_{2,{\rm P}}$        $[\arcsec]$ & $   -0.15_{- 0.04}^{+ 0.04}$ &  $ -0.19_{-0.02}^{+0.01}$  \\
 $\phantom{ }q_{\rm P\phantom{,1}}$                 & $   \phantom{-}0.78_{- 0.14}^{+ 0.11}$ &  $ \phantom{-}0.66_{-0.06}^{+0.05}$  \\
 $\phantom{ }\phi_{\rm P\phantom{,1}}$  $[\degr]$   & $  88_{-34}^{+ 10}$ &  $ 98_{-3}^{+2}$  \\
 $\phantom{ }b_{\rm P\phantom{,1}}$     $[\arcsec]$ & $   \phantom{-}5.32_{- 0.12}^{+ 0.17}$ &  $ \phantom{-}5.47_{-0.07}^{+0.10}$  \\
 $\phantom{ }\theta_{1,{\rm S}}$        $[\arcsec]$ & $  \phantom{-}5.23_{- 0.39}^{+ 0.32}$ &  $\phantom{-}5.22_{-0.11}^{+0.15}$  \\
 $\phantom{ }\theta_{2,{\rm S}}$        $[\arcsec]$ & $   -0.44_{- 0.17}^{+ 0.15}$ &  $ -0.39_{-0.04}^{+0.05}$  \\
 $\phantom{ }q_{\rm S\phantom{,1}}$                 & $   \phantom{-}0.53_{- 0.30}^{+ 0.45}$ &  $ \phantom{-}0.54_{-0.14}^{+0.15}$  \\
 $\phantom{ }\phi_{\rm S\phantom{,1}}$  $[\degr]$   & $  79_{-76}^{+73}$ &  $ 65_{-24}^{+21}$  \\
 $\phantom{ }b_{\rm S\phantom{,1}}$     $[\arcsec]$ & $   \phantom{-}0.35_{- 0.12}^{+ 0.40}$ &  $ \phantom{-}0.45_{-0.08}^{+0.16}$  \\
 $\phantom{ }t_{\rm \phantom{S,1}}$     $[\arcsec]$ & $   4.5_{- 4.0}^{+ 5.3}$ &  $ \phantom{-}1.19_{-0.39}^{+0.57}$  \\
\hline                                   
\end{tabular}
\end{table}

\subsection{Lens modelling using extended images}
\label{sec:LensModel:ExtIm}

In order to reconstruct the spatially extended surface brightness
distribution of the lensed images, we need to model both the lens mass
distribution and the extended source surface brightness distribution.
We follow \citet{SuyuEtal06} to reconstruct the source surface
brightness, denoted by $\vec{s}$, on a grid of pixels given a lens
potential model.
The reconstruction is a Bayesian regularised linear inversion, where
we denote $\lambda$ as the strength of regularisation and $\mathsf{g}$
as the form of regularisation.  This reconstruction returns the most
probable source surface brightness distribution $\vec{s}_{\rm MP}$,
and also the Bayesian evidence of source reconstruction, $\mathcal{E}$,
which is equivalent to the likelihood of the lens parameters.  The
posterior PDF for lens parameters, $\vec{\eta}$, given the extended
image surface brightness, $\vec{d}_{\rm sb}$, is
\be
\label{eq:post_esr}
P(\vec{\eta} | \vec{d}_{\rm sb}, \mathsf{g}) \propto 
\overbrace{P(\vec{d}_{\rm sb} | \vec{\eta}, \mathsf{g})}^{{\rm
    likelihood}
 } \overbrace{P(\vec{\eta})}^{\rm prior}, 
\ee
where 
\bea
\label{eq:like_esr_1}
P(\vec{d}_{\rm sb} | \vec{\eta}, \mathsf{g}) \equiv \mathcal{E} &\simeq& P(\vec{d}_{\rm
    sb} | \vec{\eta}, \hat{\lambda}, \mathsf{g}) \\
& = & \int \rm{d}\vec{s} \, P(\vec{d}_{\rm sb} |
\vec{s},\vec{\eta}) \, P(\vec{s}| \hat{\lambda}, \mathsf{g}),
\eea
and $\hat{\lambda}$ is the optimal regularisation determined by
maximising $P(\vec{d}_{\rm sb} | \vec{\eta}, {\lambda},
\mathsf{g})$.  \citet{SuyuEtal06} gave the expression for
$P(\vec{d}_{\rm sb} | \vec{\eta}, {\lambda}, \mathsf{g})$ in their
equation (19), and showed that the approximation in equation
(\ref{eq:like_esr_1}) is a valid one.

We sample the posterior of the lens parameters with MCMC methods using the
Bayesian evidence $\mathcal{E}$ of the source surface brightness reconstruction
as the likelihood at each step of the chain.  The priors on the
satellite lens parameters are as before (described in Section
\ref{sec:LensModel:profiles}).
In the sampling process,
we use $30\times30$ source pixels; the typical pixel size is
$\sim$$0.035\arcsec$. 
We find that our results
are robust under changes in the number of source pixels, provided
the source resolution is sufficient to describe the data 
($\gtrsim 20\times20$).  We assume the curvature form of
regularisation for the source surface brightness distribution
\citep{SuyuEtal06}, which is applicable in this 
case given the relatively smooth surface brightness of the lensed
images.  Furthermore, we fix the regularisation constant to be 
$\lambda_{\rm avg}$, the average of the optimal $\hat{\lambda}$ values 
from a short test chain (the optimal $\hat{\lambda}$ is different for
different lens parameter values).  In other words, we approximate equation
(\ref{eq:like_esr_1}) further and use $P(\vec{d}_{\rm sb} | \vec{\eta},
{\lambda_{\rm avg}}, \mathsf{g})$ for $\mathcal{E}$. This 
approximation is valid because 
$\mathcal{E}$ changes by $<1$ between the optimised
$\hat{\lambda}$ and the fixed $\lambda_{\rm avg}$.

\begin{figure}
\centering
\includegraphics[width=1.0\columnwidth,bb=45 170 595 705]{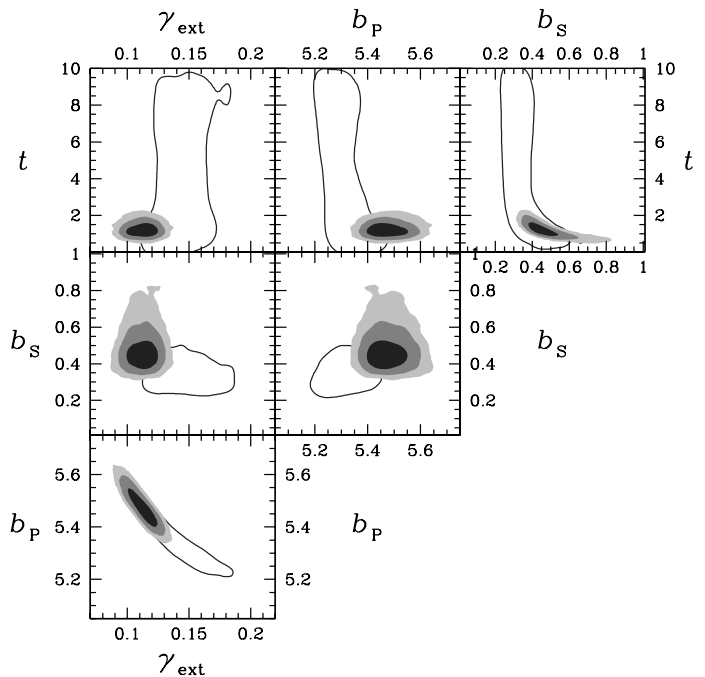}
\caption{\label{fig:EsrMCMC} Marginalised posterior PDF for
  $\gamma_{\rm ext}$ (external shear strength), $b_{\rm P}$ (strength
  of primary lens in arcseconds), $b_{\rm S}$ (strength of satellite 
  lens in arcseconds) and $t$ (truncation radius of satellite in
  arcseconds) based on extended image surface
  brightness modelling, plotted on the same scales as in 
  Fig.~\ref{fig:ImPosMCMC}.  The three shaded regions show the 68.3\%, 95.4\% 
  and 99.7\% credible regions.  For comparison, the lines show
  the 68.3\% credible regions from Fig.~\ref{fig:ImPosMCMC} with
  image positions as constraints. In contrast to the full
  $t$-degeneracy from image position modelling, extended image
  modelling allows us to constrain the truncation radius of the
  satellite galaxy halo to be  $t=\tmarg$.}
\end{figure}

The resulting constraints on the lens parameters are
shown in Fig.~\ref{fig:EsrMCMC}, plotted on the same scales as
Fig.~\ref{fig:ImPosMCMC} for comparison. The MCMC sampling converges
in $<$$10^5$ steps.
The correlations among $\gamma_{\rm
  ext}$, $b_{\rm P}$ and $q_{\rm P}$ share similar trends as in the
results of the image position modelling, but are smaller in magnitude.
In contrast to Fig.~\ref{fig:ImPosMCMC}, the strong degeneracy in $t$ is
broken with
the use of the information from the extended images.  The
significantly tighter constraints based on the extended images are in
agreement with the constraints based on image positions within the
$68.3\%$ credible regions.  Table
\ref{tab:LensParaRealData} shows the values for the marginalised lens
parameters.  In particular, the size of the satellite halo is
constrained to be $\tmarg$,
corresponding to $r_{\rm t}=\rtmarg$.

\begin{figure*}
\centering
  \includegraphics[width=2.0\columnwidth,bb=30 200 580 680]{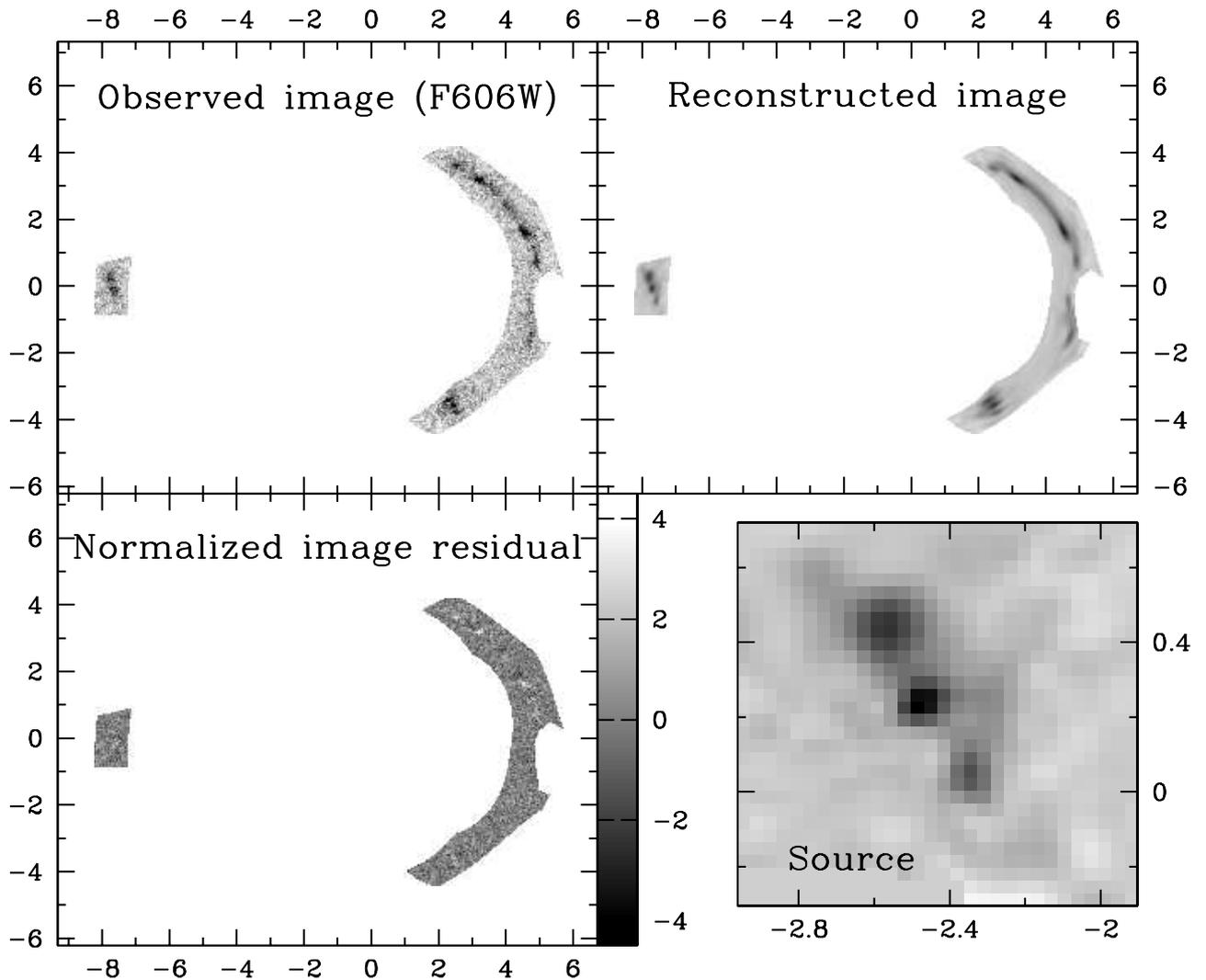}
\caption{\label{fig:EsrRec} Extended image and source surface
  brightness reconstruction.  From left to right: top: observed image,
  reconstructed image; bottom: normalised image residual (in units of
  the uncertainty estimated for each pixel), reconstructed
  source.  The morphology of the lensed arc is reproduced, and the
  reconstructed source surface brightness distribution shows three peaks.}
\end{figure*}

\begin{figure*}[ht]
\centering
\includegraphics[width=0.8\textwidth]{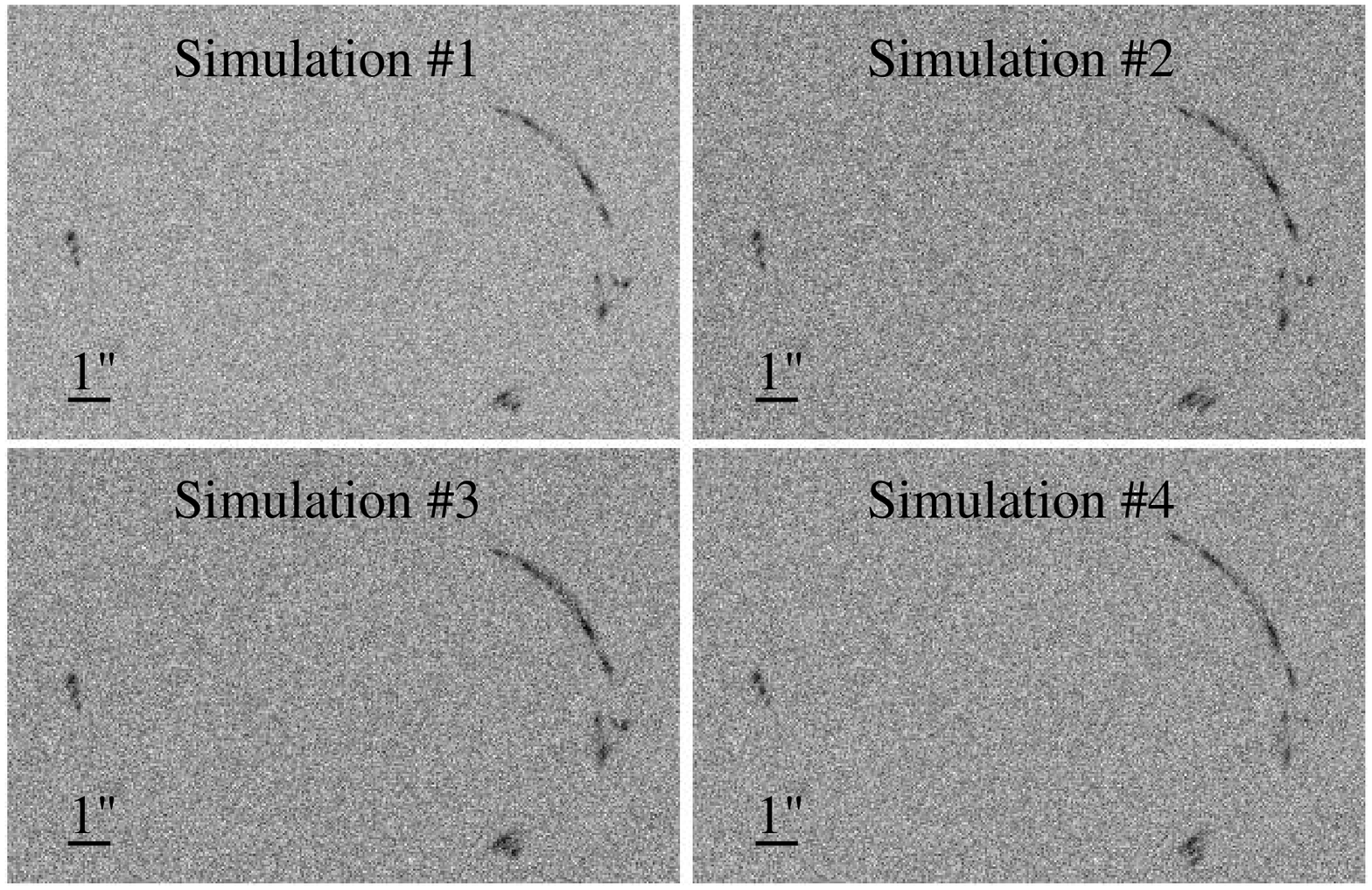}
\caption{\label{fig:SimIm} Four simulated images designed to mimic
  \ourlens\ but with different input mass distributions.  Simulations
  are labelled as \#1--4.  Details of the simulations are given in Section
  \ref{sec:sim}.} 
\end{figure*}

The most probable image and source surface
brightness reconstructions are shown in Fig.~\ref{fig:EsrRec}.  
The image pixels around the satellite
galaxy have been excluded in the modelling due to the imperfect
lens galaxy light subtraction near the galaxy centre.  One
of the strongly lensed images is also excluded (the rightmost triangle
in panel III of Fig.~\ref{fig:hstPtImMarked}) since it is blended
with the residuals of the lens galaxy light subtraction.  We opt to
discard this particular lensed image to avoid including lens galaxy light
residuals as false lensed features of the background source that could
potentially bias our modelling.  
The top and the bottom-left panels show that the lensed features,
apart from the cores of the lensed images, are reproduced by our
model.  The cores in the surface brightness clumps of galaxies
tend to be sharply peaked; this leads to the visible residuals near the cores of
highly magnified images due to source and image pixelisations.
Nonetheless, this does not limit the modelling of
the extended image structure -- the reduced $\chi^2$ is 0.75.
The reconstructed source in the
bottom-right panel indeed exhibits three peaks in its surface
brightness distribution, thus confirming our image identification in
Section \ref{sec:LensModel:ImPos}.  The critical and caustic curves of
the most probable lens parameters are illustrated in
Fig.~\ref{fig:ImPosCritCaus}.  

Comparing Tables \ref{tab:galfit} and \ref{tab:LensParaRealData}, we
see that the modelled $\phi_{\rm P}$ is offset by $\sim$$30\degr$ from
that of the observed light distribution, as was previously noted by
\citet{LimousinEtal09c}.  We attribute this to the presence of two
brightness peaks in the primary lens galaxy; in such dynamically
unrelaxed systems, it is not surprising that the observed light
distribution inside the lensed arcs does not trace the mass probed
around the arcs by strong lensing.  \mbox{N-body} simulations show that dark
matter halos can have the inner parts of their gravitational potential
elongated in a different direction from that of the outer parts
\citep[e.g.,][]{HayashiEtal07}.   Furthermore, simulations with gas indicate that
the angular momentum of the galaxy embedded in a typical dark matter
halo is on average more aligned with the inner parts than the outer
parts of the total mass distribution \citep[e.g.,][]{BettEtal10}.  
In addition, studies of intrinsic ellipticity correlations of luminous
galaxies suggest that the mass and the light of galaxies could be misaligned by
$\sim$$25\degr$$-$$35\degr$ \citep[e.g.,][]{OkumuraEtal09, BrainerdEtal09}.  The
modelled position angle of the mass distribution of the satellite is
also misaligned from that of the light by $\sim$$50\degr$, and we
discuss this in Section \ref{sec:discuss:radial_align}

The halo size measurement is obtained by modelling the F606W image, selected based on
its SNR and cosmic ray subtraction.  We could have also considered
the F475W and F814W images.  However, the F475W image has lower
SNR than the F606W image and has several cosmic rays near the lensed arcs,
which would make the analysis of F475W prone to systematic effects due to
cosmic ray residuals.  Compared to the F606W image, the F814W image
has similarly low levels of cosmic rays but lower SNR.  We repeat our
analysis on 
the F814W to check for consistency.  We find that $t$ is constrained by
the F814W data and agrees with the F606W value within the
uncertainties; however, the marginalised uncertainties of $t$ from
F814W is almost twice as large
as that of F606W, as expected due to the lower SNR of F814W. Thus, our
measurement of $t$ from F606W is consistent with F814W and would not
change drastically even if the 
F814W data were included.  In the next section, we test the robustness of this
measurement using simulations.

\section{Simulation Tests: Robustness of Halo Size Measurement}
\label{sec:simtest}

We create 4 simulations that mimic the image morphology of \ourlens.
In each case, we
have an input model which we use to lens a background source
consisting of three approximately Gaussian peaks in its surface brightness distribution,
similar to the reconstructed source in 
Fig.~\ref{fig:EsrRec}.  We then convolve the lensed image with a Gaussian
PSF of FWHM $0.08\arcsec$, and add uniform Gaussian noise.  
Figure \ref{fig:SimIm} shows the images of the four simulations labelled as 
\#1--4.  The idea is to model each of the simulations using the same
procedure as in the case of the real data to test the robustness of
the satellite halo size measurement.  The details of generating each
simulation are as follows.

\subsection{Inputs of the simulations}
\label{sec:sim}

\subsubsection{Simulation \#1}
\label{sec:sim:1}
The input mass distribution is a PIEMD for the primary lens, a dPIE
for the satellite galaxy, and a constant external shear.
The parameter $t$ for the satellite is set to $2\arcsec$, whereas the other
parameters of the input model are chosen so that the lensed image 
resembles \ourlens.  This simulation corresponds to the ideal case where
the lens profiles we use for modelling are identical to the profiles
we use to create the simulation.

\subsubsection{Simulation \#2}
\label{sec:sim:2}
This simulation is similar to Simulation \#1 (PIEMD + dPIE + constant
external shear), except the input value for the truncation radius of
the satellite galaxy is $t=5\arcsec$.  In this simulation, the input $t$ is
approximately the extent of the lensed arc.  We expect the constraints on $t$
to weaken substantially as its value becomes comparable to, or larger
than, the arc length since the morphology of the arc is not so
sensitive to a halo size that extends beyond the arc.  
This simulation is designed to probe the limit in which
we can use this method to measure the halo size of the satellite.

\subsubsection{Simulation \#3}
\label{sec:sim:3}
In this simulation, the input primary lens mass distribution is
shallower than isothermal, as one may expect for massive systems
dominated by dark matter \citep[e.g.,][]{VegettiEtal10, NewmanEtal09}.
Specifically, we describe the primary lens as a singular power-law
ellipsoid \citep[SPLE; ][]{BarKana98}, and we set the slope to be $\gamma
= 1.8$ (such that the density $\rho_{\rm 3D} \sim r^{-\gamma}$).  We
continue to use dPIE with $t=2\arcsec$ for the satellite and input a
constant external shear.  

\subsubsection{Simulation \#4}
\label{sec:sim:4}
As described in Section \ref{sec:obs} and \citet{LimousinEtal09c},
\ourlens\ is in a group environment and the external shear is mainly
generated by a concentration of galaxies located about $\sim$$1\arcmin$
to the east.  In this simulation, we create the simulated data using a
singular isothermal sphere (SIS) profile for this mass concentration
instead of a constant external shear.  The input profiles for the lens
galaxies are PIEMD and dPIE with $t=2\arcsec$.  This simulation tests
whether a constant external shear is a good approximation for the more
realistic situation where the external convergence also varies
slightly across the lens system.

\subsection{Modelling results of the simulations}
\label{sec:modsim}
\begin{figure}
\centering
\includegraphics[width=1.0\columnwidth]{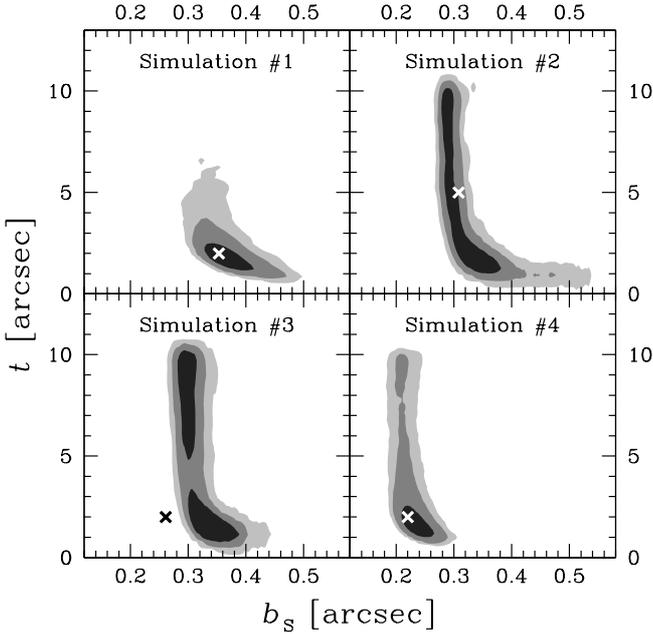}
\caption{\label{fig:sim_t_results} Marginalised posterior PDF of $b_{\rm S}$
  (strength of satellite lens) and $t$ (truncation radius of
  satellite) for the simulations.  The crosses mark the input
  parameter values.  The three shaded areas show the 68.3\%, 95.4\%
  and 99.7\% credible regions.  In the cases where the truncation
  radius is constrained (without a full degeneracy in the prior
  range), the recovered value agrees with the input value within the
  uncertainties.}
\end{figure}

We model each of the simulations using a PIEMD for the primary lens
galaxy, a dPIE for the satellite galaxy and a constant external shear,
i.e., the same profiles used to model the real data in Section 
\ref{sec:LensModel}.  We impose similar priors as for the real data on
the lens parameters: uniform with $t$ restricted to be
between $[0\arcsec,10\arcsec]$.  In each of the four simulations, the image
position modelling provides no constraints on $t$, supporting the
results in Section \ref{sec:LensModel:ImPos}.  
For modelling the extended lensed images, we employ
similar image masks as that of the real data (shown in 
Fig.~\ref{fig:EsrRec}).  Fig.~\ref{fig:sim_t_results} shows the
marginalised PDF of $b_{\rm S}$ and $t$ from modelling the extended
images in each simulation, which we describe below in turn.

In Simulation \#1, the top-left panel of Fig.~\ref{fig:sim_t_results} 
shows that the modelled truncation radius of the
satellite is $t={1.9\arcsec}_{-0.7\arcsec} ^{+1.7\arcsec}$,
which recovers the input value of $2\arcsec$.  
All the other satellite lens parameters are also recovered
within the uncertainties.  Some of the other lens parameters that
show strong degeneracies (e.g., $\gamma_{\rm ext}$, $b_{\rm P}$, and $q_{\rm P}$)
have their modelled values lie along the degeneracies but are offset
from the input values.  Nonetheless, the measurement of $t$ is not so
sensitive to these parameters since it is mostly degenerate with
$b_{\rm S}$ which is recovered within the uncertainties.  This
simulation shows that if the profiles we use to model the galaxies match
the true mass profiles, then we can recover the size of the
satellite halo within the uncertainties.  We have tested this result
using different image masks and source resolution -- in all cases,
the recovered $t$ agrees with the input value within the errors.

In Simulation \#2, the top-right panel in Fig.~\ref{fig:sim_t_results} 
shows that $t$ is now completely degenerate
with $b_{\rm S}$, and in fact with all the other lens parameters.  This panel is
based on a MCMC chain of length $10^6$ that has not yet converged -- we have not been
able to obtain a convergent chain after $>2\times10^6$ steps, in
contrast to the MCMC sampling of the other simulations that 
converges in $\lesssim$$10^6$ steps. 
Therefore, the results of this simulation suggest that 
as soon as the halo size of the satellite
becomes comparable to the arc length, even information from the
extended surface brightness does not allow the halo size to be
determined.

In Simulation \#3, there is a
mismatch in the modelled profile (PIEMD) and the input profile (SPLE)
for the primary lens galaxy. The bottom-left panel in 
Fig.~\ref{fig:sim_t_results} shows that $t$ is unconstrained within the
prior range.  The main reason for the full $t$-degeneracy is that the
model profile cannot 
accurately reproduce the simulated arcs because the relative thickness
in the extended images is sensitive to the slope
\citep[e.g.,][]{DyeWarren05, DyeEtal08, SuyuEtal10}.  The image
residuals due to the slope mismatch diminish the amount of information
on $t$ since these image residuals are more significant than residuals
caused by changes in $t$.  
Indeed, when we model the primary lens using a SPLE with a slope of
1.8 for comparison to the PIEMD model, we find that the maximum
likelihood of the SPLE model is a factor of $\sim$$e^{13}$ higher than
that of the PIEMD model.  Furthermore, we recover the input $t$
($=2\arcsec$) and $b_{\rm S}$ values for the satellite (similar to the
results of Simulation \#1) when the SPLE model is used for the primary.
Therefore, we also attribute the offset in the
recovered satellite strength $b_{\rm S}$ from the input value (shown
in Fig.~\ref{fig:sim_t_results}) to the
slope mismatch.  
We note that a formal model comparison of the SPLE and PIEMD requires
  obtaining the Bayesian evidence of the lens models (i.e.,
  marginalising over all the lens parameters), which is beyond the
  scope of this paper. 
This simulation shows that $t$ is no
longer constrained when the modelled profile is significantly
different from the input profile in terms of the slope.  This full
degeneracy was not obtained in the case of the real data, confirming
that the PIEMD+dPIE+shear model is a good description for the
\ourlens\ system.

In Simulation \#4, the modelled value of the truncation radius is
unconstrained within the $95.4\%$ CI, as shown in the bottom-right 
panel in Fig.~\ref{fig:sim_t_results}; however, within the
$68.3\%$ CI, it is constrained to be
$t=1.8\arcsec^{+0.9\arcsec}_{-0.4\arcsec}$, 
which
recovers the input value of $2\arcsec$.  Compared to Simulation \#1, the
sensitivity on $t$ is weakened due to the mismatch between the
constant external shear model and the SIS input.
Since the degeneracy is not observed in the case of the real data,
this suggests that the external convergence due to the environment
varies weaker across the strong lens system than that of an SIS (which has
$\kappa_{\rm SIS} \propto 1/\theta$), making the constant external
shear a good approximation.

The results of the simulations indicate that if the truncation radius
is constrained, then it agrees with the input value within the
uncertainties.

\section{Discussions}
\label{sec:discuss}

This is the first time the
halo size of a satellite galaxy is measured without assuming a scaling relation
and without fixing any of the satellite lens parameters based on the
observed light distribution.

\subsection {Tidal radius of the satellite}
\label{sec:discuss:tidal_radius}
Is the measured halo size of $\sim$$\rt$ for the satellite galaxy reasonable?
We can answer this question by comparing our halo size measurement to the
theoretically estimated tidal radius.

\begin{figure}
\centering
\includegraphics[width=1.0\columnwidth]{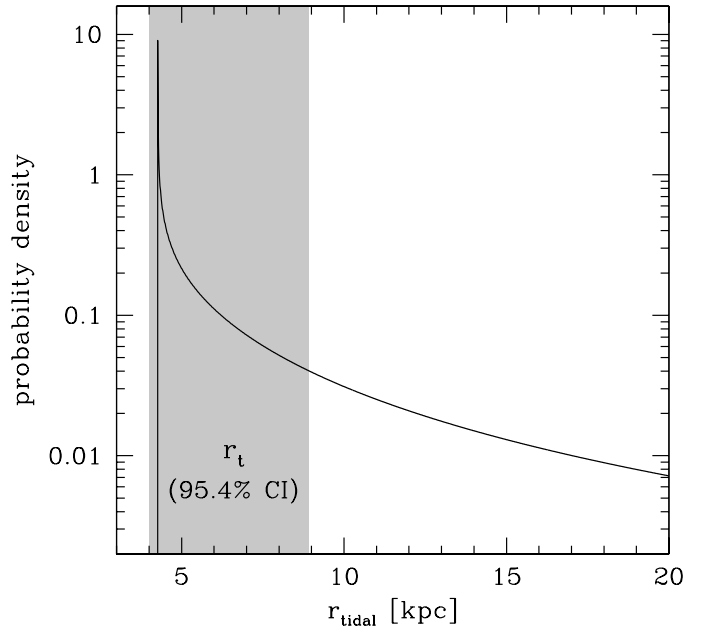}
\caption{\label{fig:P_rtidal} Probability density of the tidal
  radius of the satellite given the modelled projected distance of
  $R=25.5$\,kpc and the modelled lens strengths.  Our
  halo size measurement of $\rtmarg$ (shaded) is compatible with the
  theoretical tidal radius.}
\end{figure}

The tidal radius of the satellite is the 
location between the primary and the satellite galaxies where the
gravitational tidal force due the primary is equal to the
gravitational force of the satellite.  Mass particles
that originally belong to the satellite and are further from the
satellite centre than the tidal radius will fall into the
gravitational potential of the primary lens galaxy.  
By approximating the satellite and primary galaxies as spherical isothermal
halos, we estimate the tidal radius of the satellite halo to be
\be
\label{eq:rtidal}
r_{\rm tidal}=\sqrt{\frac{b_{\rm S}}{3b_{\rm P}}}\, r, 
\ee
where $r$ is the three-dimensional distance between the centres of the
satellite and the primary lens \citep{MetcalfMadau01}.  From
observations and lens modelling, we have only the projected
two-dimensional distance, $R$, between the satellite and the primary.
Nonetheless, we can derive the probability distribution $P(r|R)$ by
assuming that the distribution of satellites follow the total matter
distribution of the primary lens halo.  This is motivated by \mbox{N-body}
simulations that show the distribution of subhalos follow roughly the
mass distribution of the host halo \citep[e.g.,][]{NagaiKravtsov05,
  KlypinEtal10}.  To ensure that $P(r|R)$ is normalised, we impose a
hard cutoff for the isothermal distribution for the primary halo at
$r_{\rm cut}$, i.e., we assert $\rho(r)=0$ for $r>r_{\rm cut}$. For
$r_{\rm cut}$ between $500$\,kpc and $2000$\,kpc (the expected range
for a group-scale halo), we find that our PDF for $P(r|R)$ is
insensitive to $r_{\rm cut}$ for $r\ll r_{\rm cut}$.  Using the
derived $P(r|R)$, equation (\ref{eq:rtidal}), the modelled lens
strengths and the projected separation between the 
lenses of $R=5.1\arcsec=25.5$\,kpc in Table
\ref{tab:LensParaRealData},  we show in Fig.~\ref{fig:P_rtidal} the PDF for
$P(r_{\rm tidal}|R)$.  Our measurement of the truncation radius of
$\rtmarg$ based on the lens modelling in Section
\ref{sec:LensModel:ExtIm} is thus consistent with the tidal radius
PDF.

\subsection {Galaxy halo size comparison}
\label{sec:discuss:lit}

\begin{figure}
\centering
\includegraphics[width=1.0\columnwidth]{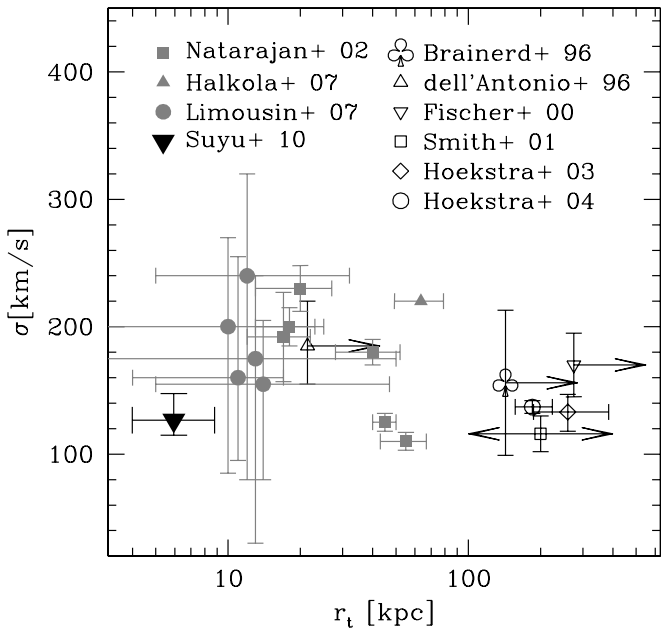}
\caption{\label{fig:tcompare} Comparison of the galaxy halo size
  measurements.  For statistical measurements, $r_{\rm t}$ is the fiducial
  truncation radius and $\sigma$ is the fiducial velocity dispersion.
  The direct measurement of our satellite
  galaxy in \ourlens\ is marked by an inverted triangle, confirming the tidal
  stripping of galaxy halos in dense environments.}
\end{figure}

We use equation (\ref{eq:sigmadpie}) to derive the fiducial velocity
dispersion of our satellite galaxy from Table
\ref{tab:LensParaRealData}: $\sigma_{\rm dPIE} = \sigmarg$. 
Figure \ref{fig:tcompare} shows our results in comparison to previous
studies of halo sizes based on strong lensing in clusters and
galaxy-galaxy lensing both in the field and in clusters.  
We note that different studies often employ different galaxy
luminosities, scaling relations and reference luminosity/velocity
dispersion, so the comparison illustrated is more qualitative than
quantitative.  Our measurement, marked by an inverted triangle, 
shows that the halo size of the satellite galaxy in
\ourlens\ is significantly smaller than isolated field galaxies of
comparable velocity dispersion.  In
fact, the satellite galaxy halo size is similar to those of galaxies in
clusters.   One may expect that galaxy members in group environments
would have individual halos which are on average larger than those in cluster
environments and smaller than isolated field galaxies.  However,
the primary lens galaxy in \ourlens\ is the brightest group galaxy
(BGG), so the satellite galaxy is probably near the centre of the
group potential\footnote{more precisely, the satellite galaxy is near
  one of the possibly two centres of the group potential since the
  luminosity map of \ourlens\ is bimodal with one peak at the BGG and
  another peak at a collection of galaxies that are $\sim$$1\arcmin$ 
  east of the BGG \citep{LimousinEtal09c}} which could make the
stripping of the satellite galaxy more efficient.  Indeed,
hydrodynamical \mbox{N-body} simulations done by \citet{LimousinEtal09b}
and the analysis of the galaxy cluster Cl 0024+16 by
\citet{NatarajanEtal09} both indicate 
that in galaxy clusters, the halo size of a galaxy is smaller
the closer the galaxy is to the cluster centre.   Our measurement of
the satellite halo size therefore supports the tidal stripping of
galaxy halos in dense environments found in earlier lensing studies.

\subsection {Radial alignment of the satellite}
\label{sec:discuss:radial_align}

As mentioned in Section \ref{sec:LensModel:ExtIm}, the total mass
distribution of the satellite is misaligned with the observed light
distribution by $\sim$$50\degr$.  Specifically, with respect to the line
connecting the modelled centres of the satellite and the primary lens
galaxies, the major axis of the
modelled satellite mass distribution is at an angle of
$49\degr^{+21\degr}_{-24\degr}$ whereas the major axis of the light distribution
is at $99\degr\pm2\degr$ ($95.4\%$ CI), where both angles are measured
counterclockwise.  Compared to the light
distribution of the satellite, its mass distribution
is elongated more towards the centre of the primary,
exhibiting the so-called ``radial alignment''.

Studies based on both observations \citep[e.g.,][]{PereiraKuhn05,
 AgustssonBrainerd06, FaltenbacherEtal07} and \mbox{N-body} simulations
\citep[e.g.,][]{PereiraEtal08, KnebeEtal08a} show that the major axes
of satellite galaxies (subhalos) tend to point towards the centre of
mass of their host.   These radial alignments are detected by averaging
over many satellite galaxies.  In the case of observations, the position angles
of the luminous parts of the satellites are measured and compared to
the direction 
of the host galaxy.  In the case of \mbox{N-body} simulations,
mass particles belonging to subhalos are identified, and the shapes of
the subhalos are measured.  By following the evolution of the alignment of
the subhalos with time, \citet{PereiraEtal08} argued for tidal
torquing as the mechanism for producing the radial alignment.  One
would expect that dark matter halos, being more extended and having
lower spins than the galaxies, would be more easily torqued.  Indeed,
\citet{KnebeEtal08b} found that the central parts of the subhalos were less
radially aligned than the subhalos as a whole. Furthermore,
\citet{PereiraEtal08} found 
that the dark matter alignment was stronger than the galaxy alignment.
Our measurement of the orientations of \textit{both} the luminous and
the total matter distributions of the satellite galaxy in
\ourlens\ also hints at the stronger total matter alignment 
that was found for ensembles of satellites in earlier studies.

\subsection{Implications for modified Newtonian dynamics}
\label{sec:discuss:MOND}

Modified Newtonian dynamics (MOND) was first suggested by
\citet{Milgrom83} as an alternative to the existence of dark matter
for explaining the flat rotation curves observed in spiral galaxies.  Instead
of requiring additional dark matter in the outer parts of a galaxy,
MOND postulates that Newtonian gravity breaks down in this
small-acceleration region, and the acceleration experienced by a test
particle is larger than the Newtonian value.  For a
review on the subject, we refer the reader to \citet{SandersMcGaugh02}.
\citet{SellwoodKosowsky01} pointed out several tests that could rule
out simple theories of modified gravity, one of which is the
misalignment between the axes of dark matter halos and those of the
visible matter (i.e., mass must follow light in MOND since there is no
dark matter).

In our analysis of \ourlens, we find that the dimensionless surface
mass density of the satellite galaxy, $\kappa_{\rm S}$, is misaligned
with that of the light by $\sim$$50\degr$.  Under General
Relativity (GR), $\kappa_{\rm S}$ is the surface mass density of the
total mass distribution, up to a constant factor.  However, in MOND, the
derived $\kappa_{\rm S}$ does not in general scale linearly with the
surface mass density \citep[e.g.,][]{ZhaoEtal06}.  The key question for
testing MOND is whether the mass distribution of the satellite galaxy
that is modelled using MOND would also be misaligned with the observed
light distribution.  \citet{MortlockTurner01} considered gravitational
lensing in MOND under the assumption that the deflection of photons is
twice that of massive particles moving at the speed of light, in
accordance with GR.  This assumption must be valid in the Newtonian
limit of any MONDian lensing theory.  The authors find that while
gravitational lensing deflections depend only on $\kappa$ in GR
(thin-lens approximation), this is no longer true in MOND where in
general the deflections depend on the mass distribution of the lens 
along the line of sight.  Nonetheless, \citet{MortlockTurner01} showed
that for a typical elliptical galaxy, the thin-lens approximation holds;
the deflection angles are the same as the GR predictions in the
inner parts (where the projected distances are $\lesssim$$2\,{\rm kpc}$)
and approach an asymptotic value in the outer parts.
\citet{ChiuEtal06} found qualitatively similar trends for the
deflection angles in a relativistic formulation of MOND by
\citeauthor{Bekenstein04} (\citeyear{Bekenstein04}; the
tensor-vector-scalar theory).  Therefore, 
modelling the satellite galaxy in MOND can be seen as modelling the
satellite galaxy in GR with a different deflection angle field, or
equivalently, a different mass radial profile.  In a strongly lensed
system, the position angle of an elliptical mass distribution is
determined based on the observed image configuration and is quite
insensitive to the actual radial profile of the mass distribution.
Therefore, even if the satellite galaxy were modelled using MOND, it
would still need to have its mass distribution elongated in a
direction that is misaligned with the light by $\sim 50\degr$ to fit
the lensing observations.  This misalignment cannot be explained by
MOND and supports the dark matter hypothesis.

\section{Conclusions}
\label{sec:conclude}

We modelled the mass distributions of the group-scale strong lens
system \ourlens\ that consists of a massive elliptical lens and a satellite lens
galaxy.  The lens galaxies are modelled as parametrised profiles, with
the truncation radius of the satellite as one parameter to
characterise its halo size.  Using Markov chain Monte Carlo methods to
sample the lens mass parameters, we compare the results based on image
position modelling and based on extended surface brightness modelling.
From this we conclude:

\begin{itemize}
\item In comparison to image position modelling, extended surface brightness modelling provides significantly
  tighter constraints on the lens parameters, although the results of the two approaches agree within
  the uncertainties.
\item Image position modelling provides no constraints on the halo size of the
  satellite galaxy. 
\item Extended surface brightness modelling constrains the halo
  size of the satellite to be $\rtmarg$ (95.4\% CI).
\item Modelling of simulations that resemble the system \ourlens\ shows
  the recovery of the input truncation radius if it is constrained.  Our
  three-component lens model (consisting of PIEMD, dPIE and constant
  external shear) provides an adequate description of \ourlens.
\item This technique works on galaxy group- or cluster-scale
  lenses where the galaxy member halo size is spatially less extended 
  than features of the lensed arcs.
\item The measured satellite halo size is in agreement with its
  estimated tidal radius, indicating that galaxies in group
  environments experience tidal stripping of their halos.  This
  confirms the tidal stripping of galaxy halos in dense environments
  found in numerical simulations and in previous lensing analysis of
  galaxy clusters.
\item The position angles of the total matter distribution and the
  light distribution of the satellite are misaligned by
  $\sim$$50\degr$. The major axis of the total matter distribution points
  more towards the centre of the primary halo, consistent with the
  detected radial alignment of satellite galaxies in previous
  observational studies and numerical simulations.
\item The misalignment between mass and light of the satellite galaxy
  is a serious challenge to MOND.

\end{itemize}

The method developed in this paper for measuring the galaxy halo sizes is
general and applicable to other lens systems.  Indeed, there are already
several group- and cluster-scale lens systems known to contain strong lensing
arcs with nearby group or cluster galaxy member(s).  This new galactic halo
ruler, when applied to a sample of systems, should help advance our
understanding of galaxy evolution and of the growth and assembly of
galaxy groups and clusters. 

\begin{acknowledgements}
  We thank P.~Schneider, R.~Cabanac, M.~Limousin, P.~Marshall, 
  J.~Chen, J.~Hartlap, P.~Behroozi for useful discussions, and 
  Y.~Mellier for helpful suggestions.  We are
  grateful to R.~Cabanac and R.~Mu\~{n}oz for
  providing us with information regarding the lens redshifts, and thank
  J.~Richard and T.~Treu for the previously published redshifts from 
  Keck/LRIS observations.  We thank the anonymous referee for helpful
  comments on the manuscript.  S.H.S.~is supported in
  part through the Deutsche Forschungsgemeinschaft under the project
  SCHN 342/7--1.  A.H.~acknowledges support from the Finnish Academy
  of Science and Letters.
  This research was supported in part by the DFG cluster of
  excellence ``Origin and Structure of the Universe''.  
  Based in part on observations made with the NASA/ESA
  \textit{Hubble Space Telescope}, obtained at the Space Telescope
  Science Institute, which is operated by the Association of
  Universities for Research in Astronomy, Inc., under NASA contract
  NAS 5-26555. These observations are associated with program 10876.
\end{acknowledgements}

\bibliographystyle{aa}
\bibliography{J0854halosize}

\end{document}